\documentclass[aps, pra, notitlepage, citeautoscript, superscriptaddress, eqsecnum, reprint]{revtex4-2}

\usepackage{xr}
\usepackage{amsmath}
\usepackage{amsthm}
\usepackage{amsfonts, amssymb,amsxtra}
\usepackage[]{graphicx}
\usepackage{grffile}
\usepackage{amsfonts}
\usepackage{framed}
\usepackage{bbm}
\usepackage{braket}
\usepackage{xcolor}
\usepackage{mathtools}
\usepackage{LatexCommands}
\usepackage{booktabs}

\usepackage{fancyhdr}
\newcommand{\nocontentsline}[3]{}
\newcommand{\tocless}[2]{\bgroup\let\addcontentsline=\nocontentsline#1{#2}\egroup}

\usepackage[colorlinks=true]{hyperref}
\hypersetup{
    unicode=false,          
    pdftoolbar=true,        
    pdfmenubar=true,        
    pdffitwindow=false,     
    pdfstartview={FitH},    
    pdftitle={},    
    pdfauthor={},     
    pdfsubject={},   
    pdfcreator={},   
    pdfproducer={}, 
    pdfkeywords={} {} {}, 
    pdfnewwindow=true,      
    colorlinks=true,       
    linkcolor=blue, 
    citecolor=blue,        
    filecolor=blue,      
    urlcolor=blue           
}

\begin{document}

\title{Mitigating Non-Markovian and Coherent Errors Using Quantum Process Tomography of Proxy States}

\author{I-Chi Chen}
\affiliation{Department of Physics and Astronomy, Iowa State University, Ames, IA 50011, USA}
\affiliation{Theoretical Division, Los Alamos National Laboratory, Los Alamos, NM 87545, USA}

\author{Bharath Hebbe Madhusudhana }

\affiliation{MPA-Quantum, Los Alamos National Laboratory, Los Alamos, NM 87544, USA}

\begin{abstract}

Detecting mitigating and correcting errors in quantum control is among the most pertinent contemporary problems in quantum technologies. We consider three of the most common bosonic error correction codes --- the CLY, binomial and dual rail and compare their performance under typical errors in bosonic systems. We find that the dual rail code shows the best performance. We also develop a new technique for error mitigation in quantum control. We consider a quantum system with large Hilbert space dimension, e.g., a qudit or a multi-qubit system and construct two $2- $ dimensional subspaces ---  a code space, $\mathcal C = \text{span}\{|\bar{0}\rangle, |\bar{1}\rangle\}$ where the logical qubit is encoded and a ``proxy'' space $\mathcal P = \text{span}\{|\bar{0}'\rangle, |\bar{1}'\rangle\}$.  While the qubit (i.e., $\mathcal C$) can be a part of a quantum circuit, the proxy (i.e., $\mathcal P$) remains idle. In the absence of errors, the quantum state of the proxy qubit does not evolve in time.  If $\mathcal E$ is an error channel acting on the full system,  we consider its projections on $\mathcal C$ and $\mathcal P$ represented as pauli transfer matrices $T_{\mathcal E}$ and $T'_{\mathcal E}$ respectively.  Under reasonable assumptions regarding the origin of the errors,  $T_{\mathcal E}$ can be inferred from $T'_{\mathcal E}$ acting on the proxy qubit and the latter can be measured without affecting the qubit.  The latter can be measured while the qubit is a part of a quantum circuit because, one can perform simultaneous measurements on the logical and the proxy qubits. We use numerical data to learn an \textit{affine map} $\phi$ such that $T_{\mathcal E} \approx \phi(T'_{\mathcal E})$.  We also show that the inversion of a suitable proxy space's logical pauli transfer matrix can effectively mitigate the noise on the two modes bosonic system or two qudits system.
\end{abstract}
\date{\today}

\maketitle


\tocless\section{Introduction}

Quantum entanglement is a central resource in most quantum technologies including computation, simulation, sensing and communication that enables quantum advantage. However, the presence of entanglement also makes quantum information fragile --- entangled states are more susceptible to errors such as decoherence and decay. Therefore, one of the central challenges in quantum information technology today is to detect, mitigate and/or correct errors. Errors in quantum control can be classified into three categories: (i) Coherent errors which can be described by a unitary operator, typically caused by miscalibration in the experimental control parameters, (ii) Markovian errors which can be described by a Lindblad master equation, caused by coupling to external environments and (iii) non-Markovian errors which encompass all other errors, caused typically by fluctuation of the external parameters~\cite{PhysRevLett.101.150402}. Most error correction protocols are targeted towards and are quite effective against Markovian errors. We show that our techniques are effective to detect and mitigate coherent and non-Markovian errors.

The problem of characterizing errors has been addressed using multiple techniques. Quantum process tomography and gate-set tomography~\cite{Nielsen_2021} (GST) provide a detailed charaterization of the errors, enabling a study of the source of these errors in the experiment~\cite{PhysRevA.107.042611, PhysRevA.105.022437}. Process tomography protocols cannot be scaled to larger number of qubits. However, one can trade the detail for scalability.  Randomized benchmarking and its variants~\cite{PhysRevLett.106.180504,  Proctor_2021} focus on a single measure --- fidelity, and are able to scale to larger systems. 
\begin{figure}[ht]
     \includegraphics[width=0.48\textwidth]{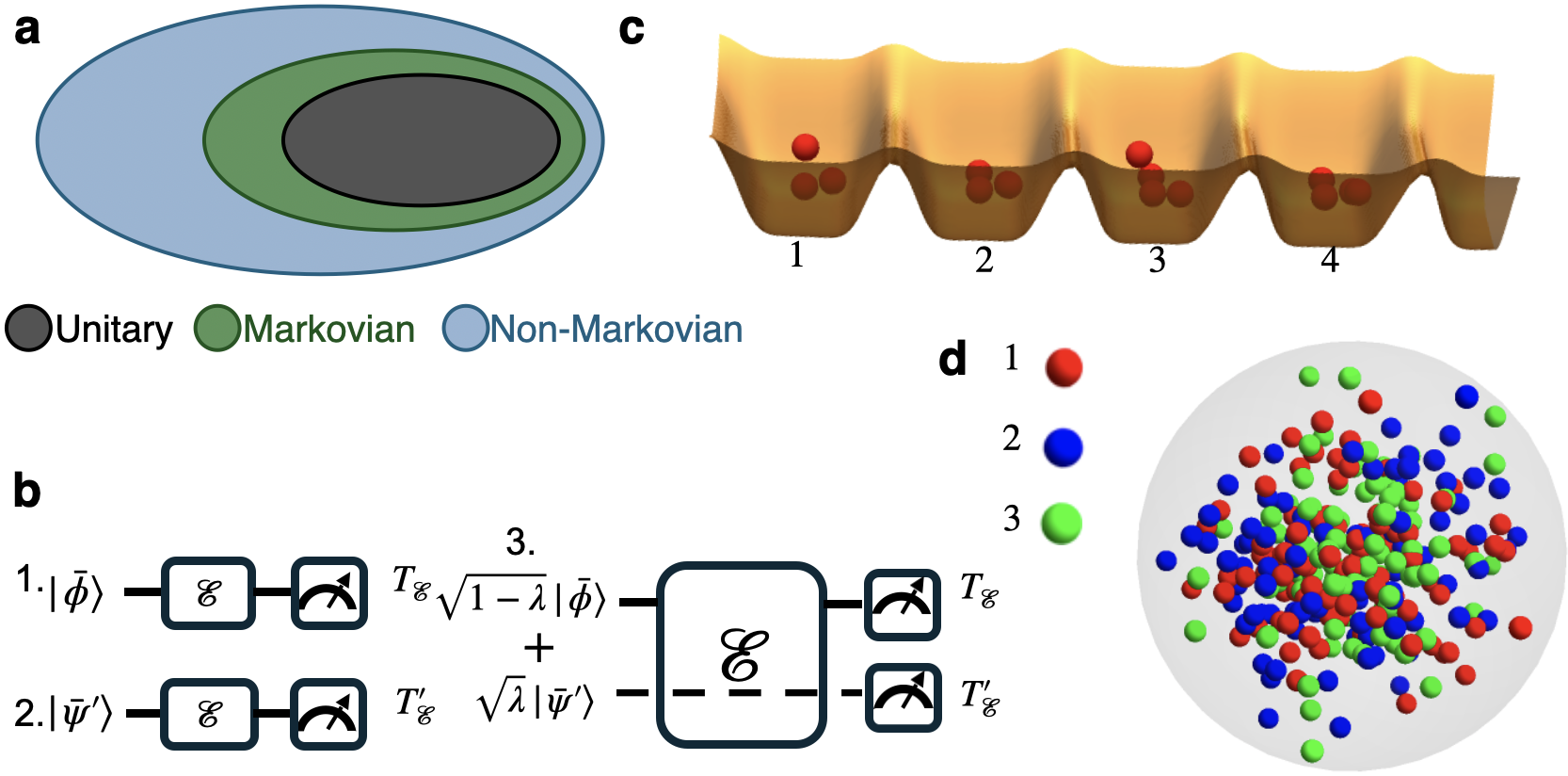}
     \caption{ \textbf{a. } Realistic errors in quantum control can be classified based on their physical origin into Coherent, Makovian and non-Markovian errors. \textbf{b.} the idea of using proxy qubits: a $d$ dimensional quantum system can be used to encode a proxy state(1) and a code state, i.e., the qubit (2). A  superposition between them can be used to perform a computation using the qubit while simultaneously characterizing the noise $\mathcal E$.  $T_{\mathcal E}$ and $T_{\mathcal E}'$ are the pauli transfer matrices corresponding to the projection of $\mathcal E$ onto the code and proxy spaces respectively.   We numerically show that $T_{\mathcal E}$ can be learnt from $T'_{\mathcal E}$. We consider a system of multi-mode bosons to embed $\mathcal C$ and $\mathcal P$.  Examples of multi-mode bosons include \textbf{c.} bosons in an optical lattice with $L=4$ sites and \textbf{d.} spinor Bose-Einstein Condensate.  
     }
     \label{fig:demo}
 \end{figure}

Error mitigation is a technique of post processing the measured data~\cite{RevModPhys.95.045005}. Probabilistic error cancellation~\cite{Temme2017,vanddenberg2023probabilistic} and zero noise extrapolation~\cite{Tiron20}. They are usually used for qunatum many body simulation on the noisy intermediate scale quantum device~\cite{kim2023utility,Chen23}. However, quantum error mitigation's limitations have also been extensively studied~\cite{quek2024exponentially}. Quantum error correction, on the other hand, uses a redundancy in the number of qubits, i.e., encodes a logical qubit in a larger number of physical qubits and uses the extra degrees of freedom to detect and correct the errors in the logical qubit It is a very well developed and an active area of research~\cite{campbell2024quantum}. While quantum error mitigation does not have a significant overhead in the number of qubit, it has strong limitations. Error correction, while being very effective, often comes with extremely high overheads in terms of the number of physical qubits per logical qubit. For the bosonic system, a logical qubit could be encoded into highe occupation number or continuous spectrum. The corresponding error correction code is known as bosonic code~\cite{albert2022}. 

The most known bosonic code is Gottesman-Kitaev-Preskill (GKP) code~ \cite{Gottesman2001,Harrington2001}, which encode a logical qubit into continuous spectrum. Nonetheless, some of bosonic system only have finite mode and finite occupation number. It's hard to implement GKP code on such system. Instead, binomial code~\cite{Michael2016}, Chuang-Leung-Yamamoto (CLY) code \cite{Chuang1997} and spin cat code~\cite{Wei2021,Omanakuttan2024} can be easily implemented on this kind of system.
Nevertheless, our result show that it's hard for the bosonic code mentioned above to get rid of the non-Markovian noise and coherent noise. Hence, here we develop an method mix quantum error correction and error mitigation. We develop a technique, where we use limited overheads to construct proxy states, measurements of which can used to re-construct  the error operation using a quantum process tomography and mitigate the errors on the logical states.  

\tocless\section{The idea and its mathematical structure}
\label{sub: Ana}

Errors in quantum control may occur during an idle time, due to unintended couplings between the qubits and the environment or during gate application due to imperfections in the gates.  In either of the cases, we may assume that the origin of the errors is physically natural, i.e., it is an \textit{undesigned } physical process that causes the error. This observation leads us to the following idea: Let us assume that the qubit is a two-dimensional space embedded inside a quantum system with additional degrees of freedom, i.e., an inherently higher dimensional Hilbert space. The errors potentially affect the entire physical system including the additional degrees of freedom, which may be used to characterize them.  In this section, we will develop a mathematical structure for this idea.

Let $\mathcal H$ be a $d-$dimensional Hilbert space of a qudit with $d\geq 4$. We identify $\mathcal C =\text{span}\{\ket{\bar{0}}, \ket{\bar{1}}\} \subset \mathcal H$ as the code space and an orthogonal space $\mathcal P =\text{span}\{\ket{\bar{0}'}, \ket{\bar{1}'}\} \subset \mathcal H$ as the proxy space. The idea is to use measurements on $\mathcal P$ to characterize errors in $\mathcal C$.  We provide one example before proceeding further. Let us consider a system of $N\geq 4$ two mode bosons with $\mathcal H = \text{span}\{\ket{0, N}, \ket{1, N-1}, \cdots, \ket{N, 0}\}$ where $\ket{i, j}$ represents a state with $i$ atoms in the first mode and $j$ atoms in the second.  We can then consider $\ket{\bar{0}}=\ket{N, 0}$, $\ket{\bar{1}}=\ket{0, N}$ and $\ket{\bar{0}'}=\ket{N-k,k}$, $\ket{\bar{1}'}=\ket{k, N-k}$ for $k>0$.

We work in the space $\mathcal C \oplus \mathcal P  \subset \mathcal H$. A normalized state $\ket{\psi}$ in this space can be written as 
\begin{equation}
    \ket{\psi} = \sqrt{1-\lambda}\ket{\bar{\phi}} + \sqrt{\lambda}\ket{\bar{\psi}'}
\end{equation}
where $\ket{\bar{\phi}}\in \mathcal C$ and $\ket{\bar{\psi}'}\in \mathcal P$ are both normalized and $\lambda > 0$. A unitary gate is constructed such that it leaves $\mathcal P$ invariant. That is
\begin{equation}
    \ket{\psi} \mapsto U \ket{\psi} = U\sqrt{1-\lambda}\ket{\bar{\phi}} + \sqrt{\lambda}\ket{\bar{\psi}'}
\end{equation}
However, in the presence of errors:
\begin{equation}\label{error_U}
    \ket{\psi} \mapsto \mathcal E (U\sqrt{1-\lambda}\ket{\bar{\phi}} + \sqrt{\lambda}\ket{\bar{\psi}'} )
\end{equation}
The errors don't leave $\mathcal P$ invariant in general and therefore, we can use measurements on $\mathcal P$ to estimate them and then mitigate the errors on $\mathcal C$. To construct this mitigator, we need to under the nature of $\mathcal E$ when projected to $\mathcal P$ and $\mathcal C$.

While $\mathcal E$ is linear on $\mathcal H$, its projection to $\mathcal C$ and $\mathcal P$ are not linear --- they are \textit{affine maps} in general due to possible leakages outside the subspace. Typically, $\ket{\bar{\phi}} =\ket{\bar{0}}$ and $\ket{\bar{\psi}'}\in \mathcal P$ is picked from a discrete 2-design:
\begin{equation}
    \ket{\bar{\psi}'} \in S =\{\ket{\bar{\psi}_1}, \cdots, \ket{\bar{\psi}_k}\}
\end{equation}
The set $S$ is a $2-$design, i.e., 
\begin{equation}
    \frac{1}{k}\sum_j \ket{\bar{\psi}_j}\bra{\bar{\psi}_j} = \frac{1}{2}\mathbbm{1}_{\mathcal{P}} = \frac{1}{2}\Pi_{\mathcal P}
\end{equation}
Here, $\Pi_{\mathcal P}$ is the projector onto $\mathcal P$. We assume that the experiment includes $\nu = k\times \ell$ statistical repetitions of each circuit, with $\ell$ repetitions of each $\ket{\bar{\psi}_i} \in S$ in the initial state. $\ket{\bar{\phi}}$ remains the same for all $k\times \ell$ repetitions. The restriction of Eq.~\ref{error_U} to $\mathcal C$ reads
\begin{equation}
\begin{split}
    (1-\lambda )\ket{\bar{\phi}}\bra{\bar{\phi}} \mapsto& \\
    &\lambda\Pi_{\mathcal C} \mathcal E(\ket{\bar{\psi}'}\bra{\bar{\psi}'})\Pi_{\mathcal C}\\
    &+ (1-\lambda) \Pi_{\mathcal C}\mathcal E( U\ket{ \bar{\phi}}\bra{\bar{\phi}}U^{\dagger})\Pi_{\mathcal C}\\
    & +\sqrt{\lambda(1-\lambda)} \Pi_{\mathcal C}\mathcal E( U\ket{ \bar{\phi}}\bra{\bar{\psi}'})\Pi_{\mathcal C}\\
    & +\sqrt{\lambda(1-\lambda)} \Pi_{\mathcal C}\mathcal E( \ket{ \bar{\psi}'}\bra{\bar{\phi}}U^{\dagger})\Pi_{\mathcal C}\\
\end{split}
\end{equation}
Here, $\Pi_{\mathcal C}$ is the projector onto $\mathcal C$. Note that the last two terms (i.e., cross terms) vanish after averaging $\ket{\bar{\psi}'}$ over $S$. Therefore, 
\begin{equation}
\begin{split}
    (1-\lambda )\ket{\bar{\phi}}\bra{\bar{\phi}} \mapsto& \frac{\lambda}{2}\Pi_{\mathcal C} \mathcal E(\Pi_{\mathcal P})\Pi_{\mathcal C}\\
    &+ (1-\lambda)\Pi_{\mathcal C}\mathcal E( U\ket{ \bar{\phi}}\bra{\bar{\phi}}U^{\dagger})\Pi_{\mathcal C}\\
\end{split}
\end{equation}
We may re-write this map as a composition of a linear and an affine map:
\begin{equation}
      (1-\lambda )\ket{\bar{\phi}}\bra{\bar{\phi}} \mapsto (1-\lambda)f_{\mathcal E}(U\ket{ \bar{\phi}}\bra{\bar{\phi}}U^{\dagger})
\end{equation}
Here, $f_{\mathcal E}$ is an affine map defined as 
\begin{equation}
    f_{\mathcal E}(X) = L_{\mathcal E} + T_{\mathcal E}(X)
\end{equation}
The first term $L_{\mathcal E} = \Pi_{\mathcal C}\mathcal E (\Pi_{\mathcal P})\Pi_{\mathcal C} $ is the leakage error and $T_{\mathcal E}$ represents the quantum error. Note that the leakage vanishes when $\mathcal E$ is identity. 

Similarly, the restriction of Eq.~\ref{error_U} to $\mathcal P$ is an affine map:
\begin{equation}
    g_{\mathcal E}(X) = L_{\mathcal E}' + T'_{\mathcal E}(X)
\end{equation}
where
\begin{equation}
    L'_{\mathcal E} = \Pi_{\mathcal P}\mathcal E (U \ket{\bar{\phi}}\bra{\bar{\phi}}U^{\dagger}) \Pi_{\mathcal P}
\end{equation}

We can experimentally characterize $T_{\mathcal E}'$ and use it to estimate $T_{\mathcal E}$. There are two fundamental questions to be addressed before implementing this. 
\begin{itemize}
\item[i.] \textbf{Minimizing leakage:}  Do there exist $\mathcal P$  and $\mathcal C$ such that the leakages $L_{\mathcal E}$ and $L'_{\mathcal E}$ are negligible, i.e., bounded below a given threshold, for a given class of error channels $\mathcal E$?
\item [ii.] \textbf{\textit{Learnability} of errors:} Do there exist $\mathcal P$  and $\mathcal C$ such that $T_{\mathcal E}$ is \textit{learnable} from $T_{\mathcal E}'$ for a given class of error channels $\mathcal E$?
\end{itemize}
In the next sections,  we address these two questions numerically for a specific physical system --- multi-mode bosons and a standard error channel.

\tocless\section{The system and the error model}

\label{sec:two}

We consider a set of multi-mode bosons. That is, a set of $N$ bosons each of which can occupy one of $L$ states, i.e., modes. The modes can be internal spin states (e.g., spinor Bose-Einstein Condensates, realized experimentally using ultacold atoms (Fig.~\ref{fig:demo}d, ref.~\cite{RevModPhys.85.1191}) or spatially distinguishable states, such as sites in an optical lattice (Fig.~\ref{fig:demo}c),  also realized using ultra cold atoms~\cite{doi:10.1126/science.aal3837}.  We use $\hat{b}_i^{\dagger}$ to denote the modes, for $i=1, \cdots, L$.

We consider an error model with three categories of error (Fig.~\ref{fig:demo}a): (i) Unitary, (ii) Markovian and (iii) non-Markovian errors.  We model unitary errors by a \textit{residual} Hamiltonian $H$. Following the standars terms present in multi-mode bosons, we consider the following Hamiltonian
\begin{align}
    H=\sum_{i=1}^{L}\Delta_i b^{\dagger}_{i}b_{i}+J\sum_{i=1}^{L-1}b^{\dagger}_{i}b_{i+1}+U\sum_{i=1}^{L} \frac{n_i(n_i-1)}{2}
\end{align}
where $b_i$ ($b_i^{\dagger}$) is a annihilation (creation) operator for different mode $i$, and the number operator $n_i=b_i^{\dagger}b_i$, which is also know as the boson boson interaction which causes dephasing~\cite{Xiong2019}. The first term describes the energy of the mode. The second term depicts the hopping energy between the nearest neighbor modes. The third terms is the internal number particle interaction within the mode $i$.  This Hamiltonian may have a residual part as well as a fluctuating part. Fluctuation of the Hamiltonian parameters causes non-Markovian errors. To simulate this dynamics of the density matrix, we randomly pick up several the $J$ value and $\Delta_i$ value for the time evolving Hamiltonian $H_j$ based on the normal distribution with mean value $a$ and uncertainty $\sigma$ and simulate the system individually with different Hamiltonian $H_j$ and get different density matrix $\rho^f_j$ by solving the master equation. Finally, we average all of final density matrix
\begin{align}
    \rho^f=\sum^s_j \frac{\rho^f_j}{s}
\end{align}
with $s$ the number of samples.

One of the common Markovian noise in the bosonic system is the boson loss which can be depicted by the lindblad master equation
\begin{align}
    \frac{d\rho}{dt}=i[H,\rho]+\sum_i \gamma_i \left(b_i \rho b_i^\dagger -\frac{1}{2} \{b_i^\dagger b_i, \rho\} \right)
\end{align}
,where $\gamma_i$ is the decay rate of $i$th mode. Here, we consider the decay rate is time independent. As time goes by, the number of boson will exponentially decay.  See ref.~\cite{supplements} for an investigation the performance of various bosonic codes, including the $[[4,2,2,2]]$ CLY code, $0-2-4$ binomial code, and dual rail code, under this error model.\\

\tocless\section{ Proxy Code Space Logical Error Mitigation}

We use the above described system and the error model to study the relation between the errors projected into $\mathcal P$ and $\mathcal C$,  for various choices of $\mathcal P$ and $\mathcal C$ numerically.  We refer to the matrices $T_{\mathcal E}$ and $T'_{\mathcal E}$ as the logical pauli transfer matrices (LPTMs).  See ref.~\cite{supplements} for details on how the LPTM is computed.

If one can find the LPTM of logical qubit without destroying the quantum state, one can use the LPTM to do post process and correct measurement expectation value. One idea comes the experiment of the dual rydberg atom species. One of atom species is used to detect the noise, and, based on the measurement result, sequence of operations act on the another atom species to reduce the noise. Inspired by the experiment, we purpose the proxy code space which is orthogonal to the logical code space to detect the noise so one can use the corresponding measurement result to mitigate logical error for the logical qubit. However, the proxy LPTM which is obtained from the proxy code space might be different from the one obtained from logical code space. One solution is to apply the mapping from proxy LPTM to code space's LPTM. In this section, we demonstrate how to build the mapping from proxy LPTM to logical qubit's LPTM for two different cases. Moreover, we also show how to combine the proxy code space with the logical operation to mitigate the logical error.

Fig.~\ref{fig:demo} shows that the code states and proxy code states suffer from the noise. We assumed that the proxy code space's LPTM gained from the circuit of fig.~\ref{fig:demo} (b, 2) is closed to the LPTM from circuit of fig.~\ref{fig:demo} (b, 1). If so, like fig.~\ref{fig:demo} (b, 3), one can prepare the the superposition of arbitrary code states and proxy code state in logical Pauli $X$, $Y$, $Z$ basis, and then one need to simultaneously measure the the circuit to get the proxy code space's LPTM and code state's expectation value. Thus, one can inverse the noise by post-processing and get the error mitigated expectation value. If the proxy LPTM is quite different from code space's LPTM, we apply the affine mapping to transform the proxy LPTM to code space's LPTM.    
However, before the numerical experiment, we need to verify the proxy LPTM obtained from fig.~\ref{fig:demo} (b, 2) is the same as that from fig.~\ref{fig:demo} (b, 3).\\

\tocless \subsection{The choice of $\mathcal P$ and $\mathcal C$ and the leakage}

First of all, we need to investigate how to select the the proxy code space so that the noise information from the code space wouldn't affect the proxy code space. 
We take two modes bosonic system as the example. In this example, we construct code space $\mathcal{C}=\text{span}\left\{ \ket{4,8},\ket{8,4}\right\}$, of which the states is easily realized in the cold atom system.
For the proxy code states, it is required that proxy states are orthogonal to code states and behave similarly with code states. Moreover, to avoid the leakage described in subsec.~\ref{sub: Ana}, the proxy space is required to be not influenced by the code space. Thus, we choose four proxy code spaces $\mathcal{P}_{1}=\text{span}\left\{ \ket{0,12},\ket{12,0}\right\} $,  $\mathcal{P}_{2}=\text{span}\left\{ \ket{1,11},\ket{11,1}\right\} $, $\mathcal{P}_{3}=\text{span}\left\{ \ket{7,3,},\ket{3,7}\right\} $, and $\mathcal{P}_{4}=\text{span}\left\{ \ket{6,3,},\ket{3,6}\right\} $ and compare their LTPM in fig.~\ref{fig:demo} (b) and fig.~\ref{fig:demo} (c). We choose corresponding parameters $\alpha=1/\sqrt{2}$ and $\beta=1/\sqrt{2}$.

We consider two different distributions for $\Delta_i$, and $J$. One of them is that $\Delta_i$, and $J$ is randomly drawn from $\{-\sigma,\sigma\}$. Another is that $\Delta_i$, and $J$ is randomly selected from the normal distribution. For each case, we draw $1000$ samples for Hamiltonian parameter set and prepare logical $X$, $Y$, $Z$ basis states to evolve with the Hamiltonians. We also change $\sigma$ from $0.005$ to $0.02$ and see how uncertainty influences the information leakage to the proxy space. To check whether there noise information from code space leak into proxy space, we will compare the LPTM obtained from and using trace distance.   


\begin{figure}[t]
     \includegraphics[width=0.45\textwidth]{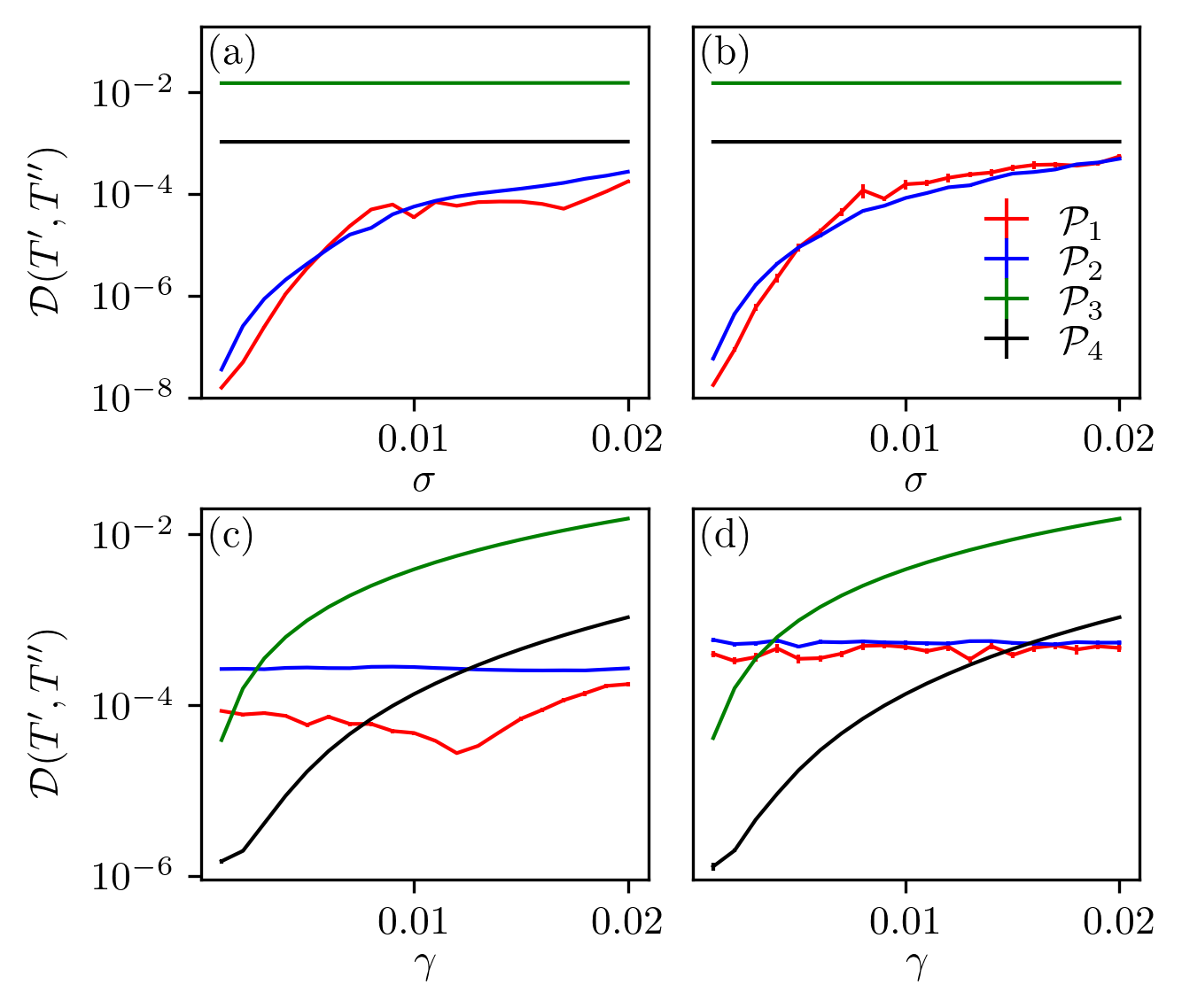}
     \caption{  The trace distance between LPTM from circuit in fig.~\ref{fig:demo} (c) and that from circuit in fig.~\ref{fig:demo} (b) varied (a,b) with $\sigma$ from $0.005$ to $0.02$ and changed (c,d) with the decay rate $\gamma$ from $0.005$ to $0.02$. The Hamiltonian parameters of leftmost panel (a,c) are drawn from distribution $\{-\sigma,\sigma\}$. That of the rightmost panel (b,d) are drawn from the normal distribution with uncertainty $\sigma$. The error bar is come from the standard error of 1000 samples.
     }
     \label{fig:compr}
 \end{figure}

\begin{figure*}[t]
     \includegraphics[width=0.85\textwidth]{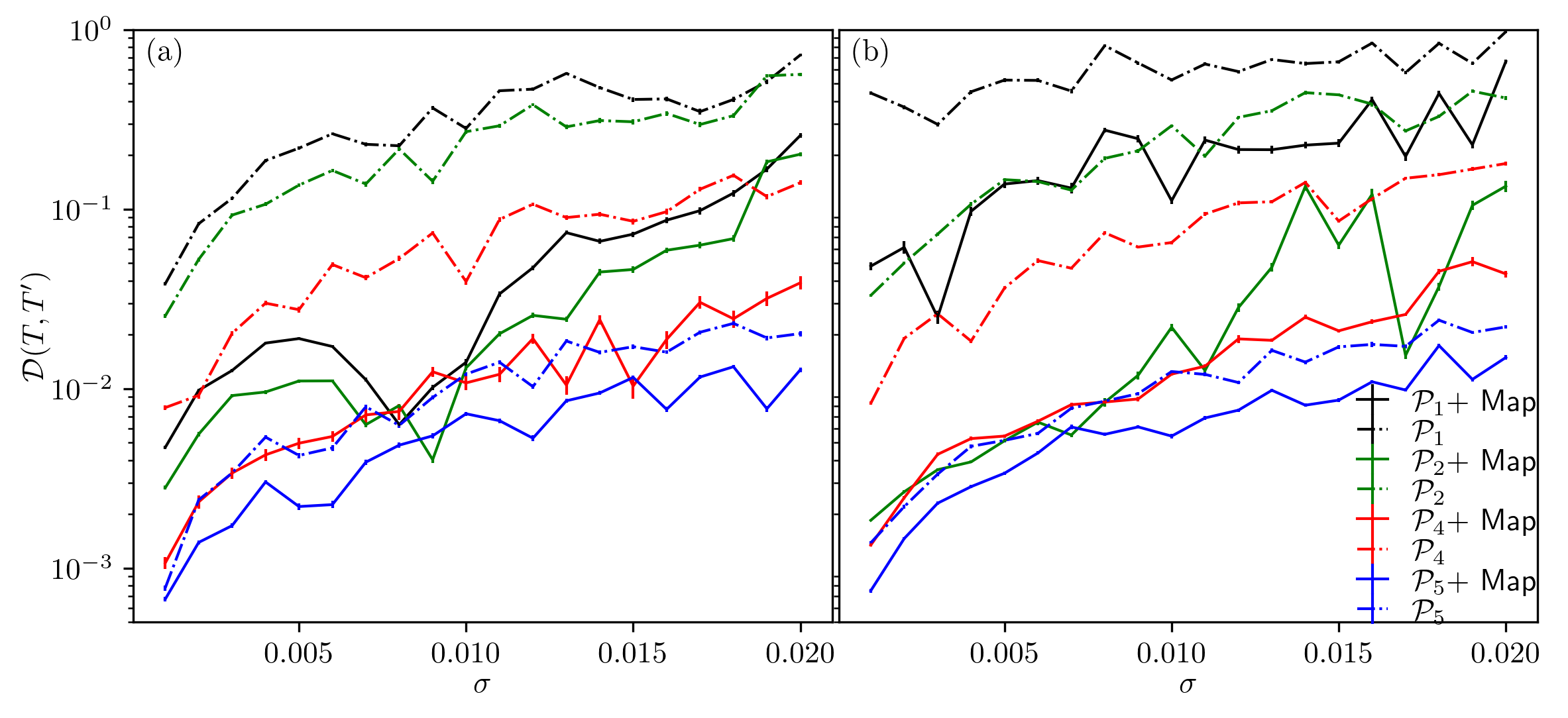}
     \caption{The trace distance between the proxy LPTM, $T'$, and real code space's LPTM, $T$, with the Hamiltonian parameters randomly selected from (a) distribution $\{-\sigma,\sigma\}$ and (b) normal distribution. The error bar is come from the standard error of 1000 samples. $\mathcal{P}_i+\text{Map}$ means the LPTM after the affine map where $i=2,3,4,5$ for different proxy spaces. On the other hand, $\mathcal{P}_i+\text{Map}$ means the LPTM without mapping.
     }
     \label{fig:P_mapping}
 \end{figure*}
 
\begin{align}
    \mathcal{D}\left(T',T''\right)=\frac{1}{2}\left\Vert T'-T''\right\Vert 
\end{align}
where $T'$ and $T''$ are LPTM from  and LPTM from  respectively with the code space measurement error detection. 

In fig.~\ref{fig:compr}, the trace distance between $T'$ and $T''$ varied with uncertainty $\sigma$ and decay rate $\gamma$ is shown. In upper panel of fig.~\ref{fig:compr}, as the uncertainty $\sigma$ increases, the trace distance of two LPTMs for $\mathcal{P}_1$ and $\mathcal{P}_2$ also increases. On the other hand, the trace distance of two LPTMs for $\mathcal{P}_3$ and $\mathcal{P}_4$ stay constant. It means that $\mathcal{P}_1$ and $\mathcal{P}_2$ suffer from information noise leakage due to the Hamiltonian uncertainty from the code space to the proxy code space but $\mathcal{P}_3$ and $\mathcal{P}_4$ don't. Since the fluctuating Hamiltonian, which we consider, includes the hopping terms that moves particles to different modes, from $\mathcal{C}$ to $\mathcal{P}_1$ requires more particle hopping than from $\mathcal{C}$ to $\mathcal{P}_2$. Thus, $\mathcal{P}_1$ suffers from leakage noise less. In contrary, in lower panel of fig.~\ref{fig:compr}, as the decay rate $\gamma$ increases, the trace distance of two LPTMs for $\mathcal{P}_3$ and $\mathcal{P}_4$ also increases but $\mathcal{P}_2$ and $\mathcal{P}_1$ are almost constant. It implies that $\mathcal{P}_3$ and $\mathcal{P}_4$ suffer from information noise leakage due to the decay.  Since only $\mathcal{P}_3$'s trace distance exceeds $0.01$, we select other proxy spaces $\mathcal{P}_1$, $\mathcal{P}_2$ and $\mathcal{P}_4$ for the further numerical simulation.     
Moreover, we also consider proxy code states of which the total excitation is more than code state's. In contrary to $\mathcal{P}_4$, we also consider $\mathcal{P}_5=\text{span}\left\{ \ket{9,6},\ket{6,9}\right\}$ for the simulation. \\

\tocless \subsection{Learnability using the affine map}
In this subsection, we test how difference between code space's LPTM (the circuit in fig.~\ref{fig:demo} (a)) and proxy code space's LPTM (the circuit in fig.~\ref{fig:demo} (b)). We also need to apply the map from proxy LPTM to the codespace's one, if proxy one is much different from codespace's. In order to find the mapping from proxy LPTM to LPTM, we construct affine map which transform the vectorized proxy LPTM to vertorized LPTM
\begin{align}
    \vec{T}=A\vec{T'}+\vec{B}
\end{align}
where $A$ is a $16 \times 16$ unknown matrix, and $\vec{B}$ is a $16$-elements unknown vector. To gain the $A$ matrix and the vector $\vec{B}$, we utilize the minimize function from Scipy package with Broyden–Fletcher–Goldfarb–Shannob (BFGS) optimization to minimize the cost function
\begin{align}
    C=\sum_{i}\left\Vert T_{i}-f\left(T_{i}'\right)\right\Vert 
\label{eq:cost}
\end{align}
where $\left\Vert .\right\Vert $ is the Frobenius norm which calculates $\left\Vert A\right\Vert =\sqrt{\text{Tr}\left[AA^{\dagger}\right]}$, and $f(.)$ is the mapping function from proxy LPTM to LPTM. 

We also consider the Hamiltonian parameters set randomly drawn from distribution $\{-\sigma,\sigma\}$ and normal distribution like the previous subsection. We randomly draw $1000$ Hamiltonian parameters set for each case and evolve logical $X$, $Y$ $Z$ basis states with those Hamiltonians. Furthermore, we also change $\sigma$ from $0.005$ to $0.02$ with totally $20$ data points to see how uncertainty affects the difference between LPTM and proxy LPTM.   
\begin{figure*}[t]
     \includegraphics[width=0.85\textwidth]{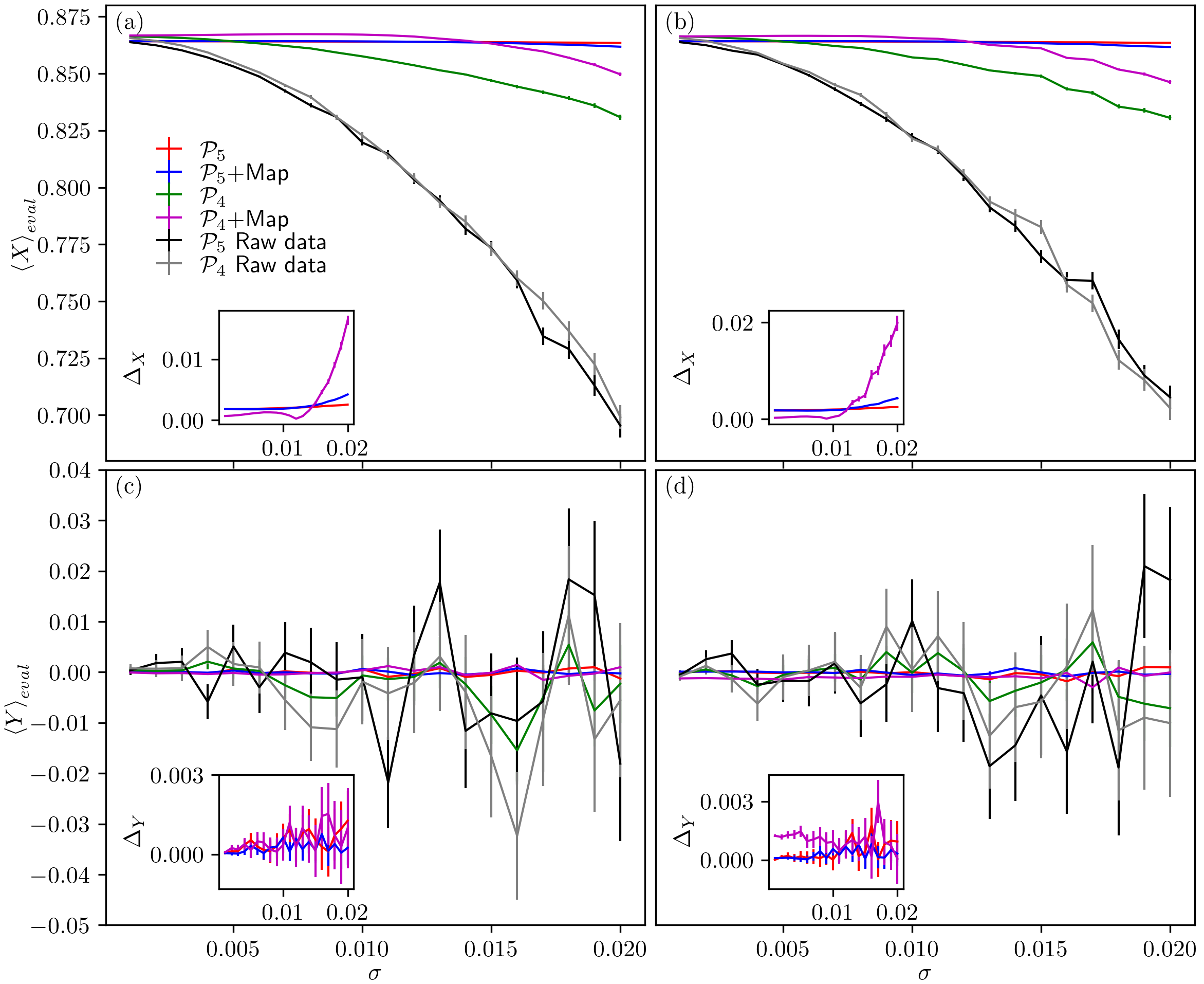}
     \caption{ The Pauli expectation values $\braket{\bar{X}}$ (upper panel) and $\braket{\bar{Y}}$ (lower panel) evaluated without error mitigation (black curve and grey curve) with error mitigation (red, blue, green, magenta curves) under the noisy Hamiltonian of which parameter set randomly drawn from (a,c) $\{-\sigma,\sigma\}$ and from (b,d) normal distribution with uncertainty $\sigma$. The inset plot shows the absolute error, defined as $\Delta_{p}=\left|\braket{p}_{eval}-\braket{p}_{exact}\right|$, between the exact expectation value, $\braket{p}_{exact}$, and the error mitigated expectation valued, $\braket{p}_{eval}$. The error bar is come from the standard error of 1000 samples.
     }
     \label{fig:eval}
 \end{figure*}
 
Fig.~\ref{fig:P_mapping} presents the trace distance between the LPTM from the proxy spaces and that from the real code space. In Fig.~\ref{fig:P_mapping} (a), with the Hamiltonian parameters are chosen from the distribution $\{-\sigma, \sigma\}$, $\mathcal{P}_1$ and $\mathcal{P}_2$'s LPTM deviates significantly from that of the code space. Although applying the affine mapping slightly improves their trace distance, both of them still remain above 0.2 at uncertainty $\sigma = 0.02$. In contrast, the other proxy spaces's LPTM is much closer than $\mathcal{P}_1$ and $\mathcal{P}_2$. Notably, $\mathcal{P}_6$ shows substantially smaller trace distances than $\mathcal{P}_4$. After applying the optimized affine mapping, the $\mathcal{P}_5$ and $\mathcal{P}_4$'s  LPTM are much closer to the real code space's. At uncertainty $\sigma = 0.02$, trace distance of $\mathcal{P}_4$ with affine map is $0.039$, and that of $\mathcal{P}_5$ with affine map is $0.015$.    
In Fig.~\ref{fig:P_mapping}(b), where the Hamiltonian parameters follow a normal distribution, $\mathcal{P}_2$ shows a slight improvement compared to $\mathcal{P}_2$ for the previous case. However, the trace distances for $\mathcal{P}_6$ remain significantly lower than that of others. At $\sigma = 0.02$, the $\mathcal{P}_4$’s trace distance with the affine map is $0.044$, which is improved by $75.7\%$ compared to that of $\mathcal{P}_4$ with out the map. On the other hand, that of $\mathcal{P}_6$ is $0.015$, which is improved by $32.4\%$.  
Overall, the affine mapping significantly reduces the trace distance between the LPTMs of the proxy spaces and the code space, bringing them closer. In the next subsection, we will explore how this technique can be applied to mitigate errors in expectation value calculations.\\

\tocless \subsection{Error mitigated expectation value}
In this subsection, we demonstrate how to mitigate logical error induced by the non-Markovian noise. We prepare initial state in fig.~\ref{fig:demo} (c) as     
\begin{align*}
    \alpha\left|\bar{\psi}\right\rangle +\beta\left|\bar{\phi}'\right\rangle 
\end{align*}
where $\left|\bar{\psi}\right\rangle$ is the target state that we want to get its expectation value and set up the target state as $\left|\bar{\psi}\right\rangle=\frac{1}{2}\left|\bar{0}\right\rangle +\frac{\sqrt{3}}{2}\left|\bar{1}\right\rangle $, and $\left|\bar{\phi}'\right\rangle$ is the state that we used it to predict LPTM of the code space from the LPTM of proxy code space. To obtain LPTM of proxy code space,  we need to prepare 4 different proxy logical states $\ket{\bar{0}'}$, $\ket{\bar{1}'}$, $\ket{\bar{+}'}$, and $\ket{\bar{Y}'_{+}}$ and measure each state's pauli $\bar{X}'$, $\bar{Y}'$, $\bar{Z}'$ expectation value after fully error detection. Meanwhile, we measure the real code state's logical Pauli $\braket{\bar{X}}$, $\braket{\bar{Y}}$, $\braket{\bar{Z}}$ expectation values with fully quantum error detection. Consequently, we gain the error unmitgated Pauli matrices expectation values $\braket{P}_{R}$ with $P$ the pauli matrix and LPTM of proxy code space $T'$. With $\braket{P}_{R}$ and $T'$, one can get error mitigated expectation value of Pauli matrices
\begin{align}
    \left(\begin{array}{c}
\braket{X}_{mit}\\
\braket{Y}_{mit}\\
\braket{Z}_{mit}\\
\end{array}\right)=T'^{-1}_{3\times3}\left(\begin{array}{c}
\braket{X}_{R}\\
\braket{Y}_{R}\\
\braket{Z}_{R}\\
\end{array}\right)
\label{eq:mit}
\end{align}
where $\braket{.}_{mit}$ means the error mitigated expectation value.

Here, we choose proxy spaces $\mathcal{P}_6$ and $\mathcal{P}_4$, which perform best based on the trace distance result in fig~\ref{fig:P_mapping}, to evaluate the $\left|\bar{\psi}\right\rangle$'s Pauli matrices expectation value. Moreover, we also change uncertainty $\sigma$ from $0.005$ to $0.02$ totally $20$ data points, and, for each data points, we randomly draw $1000$ Hamiltonian parameter sets for the simulation.

Fig.~\ref{fig:eval}, the evaluated expectation value without logical error mitigation and with logical error mitigation is presented. As the uncertainty $\sigma$ raises, the expectation values without logical error mitigation, which shown as the black curves and grey curves in fig.~\ref{fig:eval}, deviate away from the exact expectation value. The uncertainty noise makes logical error unmitigated $Y$ expectation value fluctuate a lot and $X$ expectation value decrease. However, with eq.~\ref{eq:mit}, the inversion of proxy code space's LPTM $T'$ on the noisy expectation value, the Pauli expectation value results are improved a lot. Without the affine mapping, the inversion of $\mathcal{P}_4$'s LPTM improves a bit $X$ expectation value and fluctuation of $Y$ expectation value. On the other hand, according to the inset plot of fig.~\ref{fig:eval}, that of $\mathcal{P}_5$'s LPTM improves the accuracy of the expectation value by  at uncertainty $\sigma=0.02$. With an optimized affine map, the inversion of $\mathcal{P}_4$'s LPTM further improve the accuracy of Pauli matrices expectation value, compared to that of $\mathcal{P}_4$'s LPTM without the map. However, that of $\mathcal{P}_5$'s LPTM with the optimized map doesn't further improve the accuracy of the Pauli matrices expectation values, compared to that of $\mathcal{P}_5$'s LPTM without the map. In the inset plot of fig.~\ref{fig:eval} (a), the absolute error of $\mathcal{P}_4$ with an affine map is the lowest when uncertainty is below $0.014$. When uncertainty is above $0.014$, the absolute error of $\mathcal{P}_5$ without the map becomes the lowest. In inset plot of fig.~\ref{fig:eval} (b), the absolute error of $\mathcal{P}_4$ with an affine map is also the lowest when uncertainty is below $0.012$. When uncertainty goes beyond $0.012$, that of $\mathcal{P}_5$ without the map becomes the lowest. For the pauli $Y$ expectation value, shown as the inset plots of fig.\ref{fig:eval} (c) and (d), overall, absolute error of $\mathcal{P}_5$ with the map is the lowest one. The optimized affine map can improve the expectation value result if the error mitigated result is still bit far the exact result. However, if the error mitigated result is closed to the exact result, the affine map might not improve the result.   \\


\tocless \section{ Conclusion and Outlook}
We propose a proxy code space logical error mitigation approach. Our numerical result demonstrate that LTPM from appropriate proxy code space, which has similar difference of particle number in each mode and has no leakage issue as mentioned at the beginning sec.~\ref{sec:two}, can significantly mitigate the error induced by the non-Markovian error. We also provide the numerical result showing the error mitigated expectation value with the suitable proxy code space's LPTM has just within $0.02$ absolute error compared to the exact expectation value. We also introduce the affine map to transform the proxy code space's LPTM to that of true code space. With the optimized affine mapping, the trace distance between real code space's LPTM and proxy code space's LPTM become smaller. It means that one can use proxy code space's LPTM with the map to mitigate noise. However, when original proxy space's trace distance is below $0.03$, with the map, the error mitigated expectation value might not be improved, but absolute error is still below $0.005$.

Quantum simulation is one of the areas where our technique could find applications. Unlike quantum computers, simulators are based on continuous time evolution under a constant Hamiltonian of a multi-qubit system. As a result, one of the challenges is to characterize errors~\cite{Hauke_2012}. We have shown that our technique is very effective in multi-mode multi atom bosonic systems and is therefore naturally applicable to quantum simulators. Quantum control, error characterization and mitigation of qudit systems is being explored in ultracold atomic systems~\cite{PhysRevX.14.031017, PhysRevLett.114.240401, PhysRevLett.119.150401}. Our results can be applied readily in this area as well.

Most quantum error correction codes are targeted for quantum computation applications. When the quantum advantage is exponential, it asymptotically survives the large overhead in the number of physical qubit per logical qubit. However, in quantum sensing, the advantage is at best polynomial. It is bounded by $\sqrt{N}$, $N$ being the number of qubits. Quantum sensing suffers from the same kind of errors in quantum control. While quantum error correction has been applied to quantum sensing~\cite{reiter2017dissipative, zhou2018achieving}, the large overheads often overwhelm the small ($\sqrt{N}$) quantum advantage, turning the exercise futile. Our technique, with its small overheads are therefore suitable to apply to quantum sending. This is a promising application of our technique. However, this proxy space noise mitigation technique with affine map could fail to work under some unknown noise or larger uncertainty noise. Hence, we probably need some machine learning or deep learning technique to find out suitable mapping from proxy LPTM to real code space's LPTM. We leave this exploration with machine learning techniques to the future work. Moreover, this technique also has potential application to qubit system. We also leave further exploration to our future work.\\




\tocless \acknowledgments
The authors acknowledge valuable discussions with Pei-Kai Tsai, Sivaprasad Omanakuttan. I.-C. C. and B. H. M. was supported by the U.S. DOE through a quantum computing program sponsored by the LANL Information Science $\&$ Technology Institute. Research presented in this article was supported by the Laboratory Directed Research and Development program of Los Alamos National Laboratory under project number 20230779PRD1. Portions of this work were also supported by the U.S. Department of Energy, Office of Science, National Quantum Information Science Research Centers, Quantum Science Center (analytical results presented in section II).  This work is approved for unlimited release under LA-UR-24-30653.

\bibliography{Refs}

\appendix

\cleardoublepage

\setcounter{figure}{0}
\setcounter{page}{1}
\setcounter{equation}{0}
\setcounter{section}{0}

\renewcommand{\thepage}{S\arabic{page}}
\renewcommand{\thesection}{S\arabic{section}}
\renewcommand{\theequation}{S\arabic{equation}}
\renewcommand{\thefigure}{S\arabic{figure}}
\onecolumngrid
\begin{center}
\huge{Supplementary Information}
\vspace{5mm}
\end{center}
\twocolumngrid
\normalsize
\tableofcontents

\section{Logical Error Characterization}
\begin{figure*}[t]
     \includegraphics[width=0.95\textwidth]{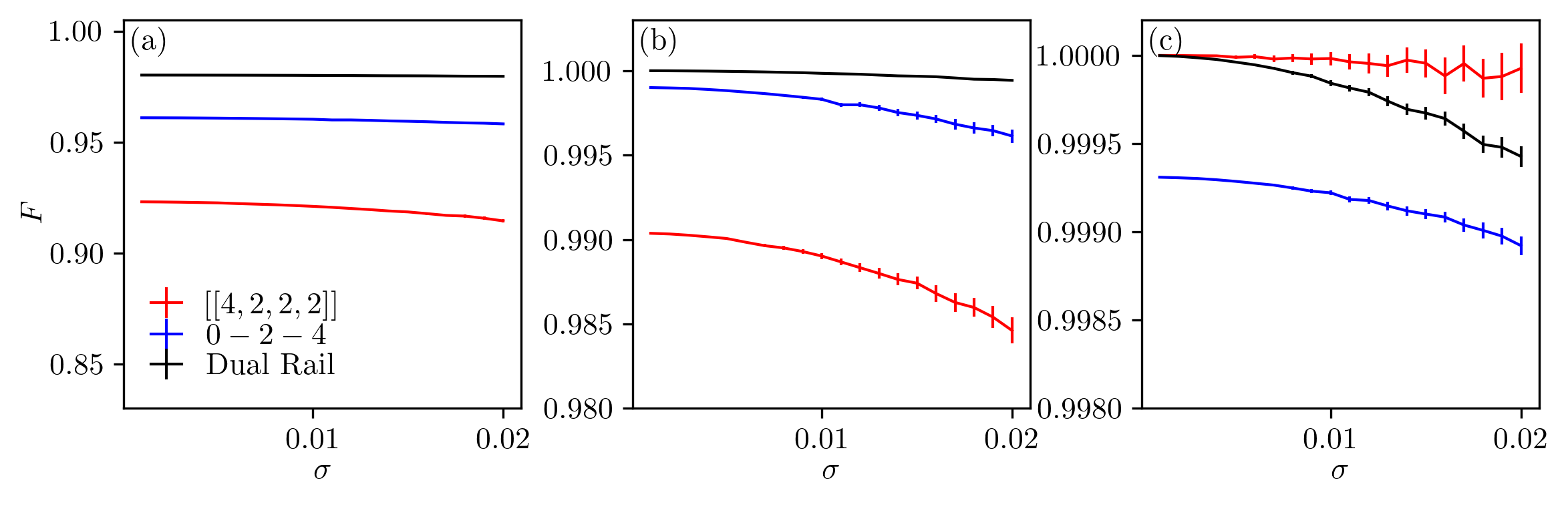}
     \caption{ The fidelity calculated using the LPTM obtained (a) without any error detection measurement, (b) with projective number measurement, and (c) with projective code space measurement. The error is calculated as the standard error of 1000 samples.   
     }
     \label{fig:Fid}
 \end{figure*}
To examine the performance of each code against the boson loss noise and the non Markovian noise, we use code space to construct the logical Pauli transfer matrix (LPTM)
\begin{align}
    T^{(R)}_{i,j}=\frac{1}{d}\textrm{Tr}\left[p_{i}\Lambda\left(p_{j}\right)\right]
\end{align}
where $p_i$ is the Pauli matrices constructed by code space and $\Lambda\left(p_{j}\right)$ is a density matrix vectorized in the Pauli matrices basis
\begin{align}
    \Lambda\left(p_{j}\right)=\sum_{i}c_{ij}\mathcal{N}\left(\left|\bar{i}\right\rangle \left\langle \bar{i}\right|\right)
\end{align}
where $\mathcal{N}$ is the noise channel including the atom loss and the non-Markovian noise, $\left|\bar{i}\right\rangle \in\left\{ \left|\bar{0}\right\rangle ,\left|\bar{1}\right\rangle ,\left|\bar{X}_{+}\right\rangle ,\left|\bar{Y}_{+}\right\rangle \right\}$ and 
\begin{align}
    c_{ij}=\left(\begin{array}{cccc}
1 & 1 & 1 & 1\\
0 & 0 & 1 & 0\\
0 & 0 & 0 & 1\\
1 & -1 & 0 & 0
\end{array}\right)
\end{align}
According to the Knill-Laflamme conditions for the quantum error detection, the error can be detected if the error makes logical state out of the code space. That is, the quantum error detection allow one to discard the errors make logical states into error space. Thus, the LPTM after projection into codespace becomes
\begin{align}
    T_{ij}^{(c)}=\frac{1}{d}\text{Tr}\left[p_{i}\Lambda_{c}\left(p_{j}\right)\right]
\end{align}
where $\Lambda_{c}$ is the code space mapping
\begin{align}
    \Lambda_{c}\left(p_{j}\right)=\sum_{i}c_{ij}\frac{P_{c}\mathcal{N}\left(\left|\bar{i}\right\rangle \left\langle \bar{i}\right|\right)P_{c}}{\text{Tr}\left[P_{c}\mathcal{N}\left(\left|\bar{i}\right\rangle \left\langle \bar{i}\right|\right)P_{c}\right]}
\end{align}
with $P_c$ the projector into the code space. However, in order to detect the error makes logical state out of code space, one need to make a projective measurement which gives code space $+1$ and error space $-1$. The corresponding measurement is hard to realize in the real hardware. Instead of implementing the projective measurement, we also consider the non-destructive total particle measurement and individual particle measurement. For instance, CLY state's excitation number is exactly $N$. If atom loss occurs, one can use non-destructive total number measurement to detect the error. For the binomial code, the code state is composed of Focks states with $N \text{mode}(S)=0$. Thus, one can use the projective measurement which gives fock states satisfying $N \text{mode}(S)=0$ condition $+1$ and others $-1$ to detect the atom loss. The corresponding LPTM can be defined as
\begin{align}
    T_{ij}^{(M)}=\frac{1}{d}\text{Tr}\left[p_{i}\Lambda_{M}\left(p_{j}\right)\right]
\end{align}
where $\Lambda_{M}$ is the code space mapping
\begin{align}
    \Lambda_{M}\left(p_{j}\right)=\sum_{i}c_{ij}\frac{P_{M}\mathcal{N}\left(\left|\bar{i}\right\rangle \left\langle \bar{i}\right|\right)P_{M}}{\text{Tr}\left[P_{M}\mathcal{N}\left(\left|\bar{i}\right\rangle \left\langle \bar{i}\right|\right)P_{M}\right]}
\end{align}
with $P_{M}$ the projector to project the incorrect number Focks states out. Since CLY $[\![ 4,2,2, 2]\!]$ code is also a two mode binomial code, one can not only do the total number measurement but also individual particle number projective measurement to detect the error. 

Once we have the LPTM, we can calculate the corresponding process fidelity of the logical qubit
\begin{align}
    F=\frac{1}{4}\textrm{Tr}\left[T_{L}T^T_{I}\right]
\end{align}
where $d$ is the Hilbert space dimension, $T_L$ is the LPTM obtained from the code states and $T_{I}$ is the ideal LPTM which is the $4 \times 4$ identity matrix here. Thus, we employ the process fidelity as a metric to compare different codes' performance. 

In the numerical experiment, we will prepare dual rail code, CLY $[\![ 4,2,2, 2]\!]$ code, and $0-2-4$ binomial code's logical X, Y, Z basis state as initial states. We implement the dynamics simulation as sec.~\ref{sec:two} described with evolving time $1$, $\gamma=0.02$, and the mean value of $\Delta_i$, $J$, $U_i$ equal to zero. We change the uncertainty of those value from $0.005$ to $0.02$ with total $20$ data points and construct the LPTM per data point.  


In fig.~\ref{fig:Fid}, the comparison of the fidelity using different codes and error detection measurements is manifested. 

To further understand the noise, we quantify the noise using LPTM. The leakage noise can be described by
\begin{align}
    L_{\mathcal{N}}=\sqrt{\left(1-T_{00}\right)^{2}+T_{01}^{2}+T_{02}^{2}+T_{03}^{2}}
\end{align}
If there is no leakage noise, the quantity is just zero. The leakage noise here describe information loss from the the logical qubit. Moreover, the Markovian noise can be depicted by 
\begin{align}
    M_{\mathcal{N}}=\sqrt{T_{10}^{2}+T_{20}^{2}+T_{30}^{2}}
\end{align}
The coherent noise and the non Markovian noise can be gained by single value decomposing the matrix $T_{3 \times 3}$
\begin{align}
    T_{3\times3}=UDV^{\dagger}
\end{align}
where $U$ and $V$ are complex unitary matrices, $D$ is the diagonal matrix, and
\begin{align}
    T_{3\times3}=\left(\begin{array}{ccc}
T_{11} & T_{12} & T_{13}\\
T_{21} & T_{22} & T_{32}\\
T_{31} & T_{32} & T_{33}.
\end{array}\right)
\end{align}
The unitary noise is described by
\begin{align}
    C_{\mathcal{N}}=\left\Vert I_{3\times3}-UV^\dagger\right\Vert 
\end{align}
where $I_{3\times3}$ is a $3 \times 3$ identity matrix. The non- Markovian noise is quantified as
\begin{align}
    N_{\mathcal{N}}=\left\Vert D\right\Vert. 
\end{align}
We fix the uncertainty $\sigma=0.02$, and see what kind of logical noise described above is induced by the uncertainty of the Hamiltonian for the individual bosonic code. 
\begin{figure}[t]
     \includegraphics[width=0.45\textwidth]{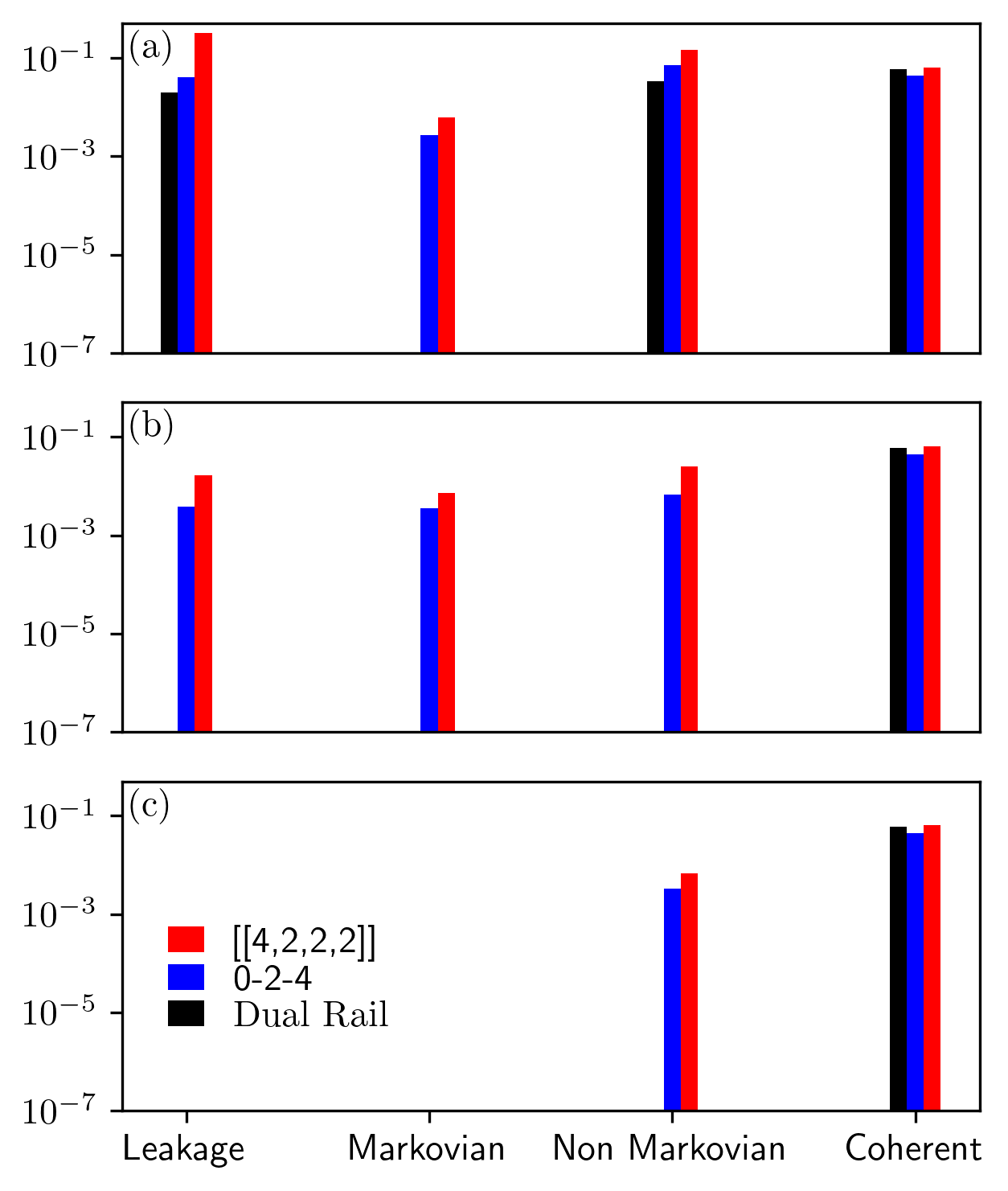}
     \caption{ The histogram of the different kinds of noise for each bosonic code (a) without any error detection measurement, (b) with projective number measurement, and (c) with projective code space measurement.
     }
     \label{fig:hist}
 \end{figure}

Fig.~\ref{fig:hist} shows the histogram of four kinds of logical noise.  In fig.~\ref{fig:hist} (a), without any quantum error detection, the logical noise is composed of leakage noise, Markovian noise, non Markovian noise, and unitary noise. For the dual rail code, the Markovian noise is surpassed without any quantum error detection. In contrast, $[\![ 4,2,2, 2]\!]$ code has the most noise among three different codes. With the number projective measurement, as fig.~\ref{fig:hist} (b) shown, the leakage noise and non-Markovian noise are significantly reduced. Particularly, the both kind of noise from dual rail code are totally ruled out. On the other hand, the unitary noise remains the same. In fig.~\ref{fig:hist} (c), with fully quantum error detection, the leakage noise is reduced to zero because the fully quantum error detection project out all of the error space. Moreover, the Markovian noise of three bosonic code is also ruled out by the fully quantum error detection. However, the coherent noise is still the same. Hence, to reduce the coherent noise, one need to use other methods like quantum
error mitigation. In next section, we propose a novel quantum logical error mitigation technique to address the coherent noise.

\section{The bosonic code}
The bosonic code is the the error correction code encoding the qubit to higher occupational state or to the continuous spectrum instead of just multi-qubit state. Gottesman-Kitaev-Preskill (GKP) code is the most common code for the bosonic system. However, since the some bosonic system like cold atom system just have finite modes and finite occupational number, the suitable code for cold atom system are binomial code and CLY code. With suitable truncation for the occupation number, the approximate cat's code, in the superposition of two coherent states with opposite sign of the displacement, also can be applied to the bosnic system with occupation number. Moreover, the coherent states can be replaced by the spin coherent states. We will discuss more detail later.

\subsection{CLY code}
CLY code encodes the $k$ qubits into $n$ modes. The codewords are in the superposition of several focks states which have $N$ number atom occupations exactly. The code protects the qubit suffering from the amplitude damping error \cite{Chuang1997}. By non-destructively measuring the particle number, one can detect the amplitude damping error and the atom loss. The corresponding notation is given by $[\![ N,n,2^k, d]\!]$ with d the minimal distance between code words.

Dual Rail code~\cite{Chung1995,Levine2024} is a kind of CLY code of which codewords composed of two bosonic mode $\{\ket{01},\ket{10}\}$. It can detect the one atom loss by the odd number parity measurement. One could make odd number parity measurement to detect the atom loss error. 

Another two mode CLY code is $[\![ 4,2,2, 2]\!]$ code. The corresponding codewords are
\begin{align}
\begin{split}
    \ket{\bar{0}}&=\frac{1}{\sqrt{2}}\left[\ket{40}+\ket{04}\right]\\
    \ket{\bar{1}}&=\ket{22} 
\label{eq:2mCLY}
\end{split}
\end{align}

which is also the two mode binomial code. Hence, one can also non-destructively measure individual mode to detect the error. If the individual mode's particle number isn't even, it means that there are errors happen to the state. In the photonic system, one could create a noon state $\ket{N}=\frac{1}{\sqrt{2}}(\ket{N,0}+\ket{0,N})$ with $N$=4 using a polarizing beamsplitter and Mach-Zehnder interferometer \cite{Afek2010}.

 \subsection{Binomial code}
Binomial code is designed to against some amount of loss and raising. One of the example is $0-2-4$ error detecting code with the codewords
\begin{align*}
    \ket{\bar{0}}&=\frac{1}{\sqrt{2}}\left[\ket{0}+\ket{4}\right]\\
    \ket{\bar{1}}&=\ket{2}
\end{align*}
The more general code protecting $G$ gain errors, $L$ loss errors, and $D$ dephasing errors is given
\begin{align*}
    \ket{\bar{0}/\bar{1}}=\frac{1}{\sqrt{2^{N}}}\sum_{p\,\textrm{even/odd}}^{\left[0,N+1\right]}\sqrt{\left(\begin{array}{c}
N+1\\
p
\end{array}\right)}\ket{p(S+1)}
\end{align*}
where the spacing $S=L+G$, the maximum order $N=\textrm{max}\left\{ L,G,2D\right\}$. The coefficients of the Focks states are including binomial coefficients so the code is also called binomial code. However, this kind of superposition state is hard to be realized in cold atom system. 

To avoid the noise like $\exp{(-\hat{n} \kappa \delta t)}$, one could extend the $1$ mode binomial code to $2$ modes binomial code. The general codeword is given by
\begin{align*}
    \ket{\bar{0}/\bar{1}}=\sum_{p\,\textrm{even/odd}}^{\left[0,N+1\right]}\frac{c_p}{2^{N/2}}\ket{p(S+1),N_{\textrm{Tot}}-p(S+1)}
\end{align*}
where $N_{\textrm{Tot}}=(N+1)(S+1)$ is the excitation number, and $c_{p}=\sqrt{\left(\begin{array}{c}
N+1\\
p
\end{array}\right)}$. For instance, when $S=1$ and $N=1$, the code state happens to be same as two mode $[\![ 4,2,2, 2]\!]$ CLY code in eq.~\ref{eq:2mCLY}.

\subsection{ Cat code}
One of the cat's code, $2$-cat's code \cite{Leghtas2013,Hastrup2022}, is in the superposition of two coherent states with opposite sign of displacement
\begin{align}
    \ket{\bar{0}/\bar{1}}=\frac{1}{N_\alpha^\pm}\left(\ket{\alpha}\pm\ket{-\alpha}\right)
\end{align}
where $N_\alpha^\pm$ is the normalized factor, and $\ket{\alpha}$ the coherent state
\begin{align}
    \ket{\alpha}=D(\alpha)\ket{0}
\end{align}
with $D(\alpha)$ the displacement operator
\begin{align}
    D\left(\alpha\right)=\exp\left[\alpha a^\dagger-\alpha^* a\right]
\end{align}
Since the coherent state is superposition of infinite Focks states, the truncation for the number of focks state is needy to apply cat's state to cold atom system. The correspond code becomes approximate $2$-cat's code. The code can surpass the dephasing noise especially when $\alpha$ is large. However, the loss noise increase linearly with $\left|\alpha\right|^{2}$ \cite{Schlegel2022}. One should find the balance between the loss noise and dephasing noise. Moreover, one can also apply the squeeze operator
\begin{align}
    S\left(\xi\right)=\exp\left[\frac{1}{2}\left(\xi^{*}a^{2}-\xi a^{\dagger2}\right)\right]
\end{align}
on the $2$- cat code so the code becomes two component cat's code of which codewords are
\begin{align}
    \ket{\bar{0}/\bar{1}}=\frac{1}{N^\pm_{\alpha,\xi}}\left(\ket{\alpha,\xi}\pm\ket{-\alpha,\xi}\right)
\end{align}
with $N^\pm_{\alpha,\xi}$ the normalized factor and $\ket{\alpha,\xi}=D\left(\alpha\right)S\left(\xi\right)\ket{0}$ the squeeze state.


Alternatively, one can replace the coherent state with spin coherent state $\ket{\theta,\phi}$. The corresponding spin 2- cat's code becomes
\begin{align}
    \ket{\bar{0}/\bar{1}}=\frac{1}{N_{\theta,\phi}}(\ket{\theta,\phi}\pm \ket{\pi-\theta,\pi-\phi})
\end{align}
where $N_{\theta,\phi}$ is the normalized factor and
\begin{align}
    \ket{\theta,\phi}=e^{-iS_{z}\theta}e^{-iS_{y}\phi}\ket{J,m=J}
\end{align}
with $\ket{J,m=J}$ is spin state with $S^2=J(J+1)$ and $S_z=J$ which is different from the focks state. The spin coherent state can also be written as
\begin{align}
    \ket{\theta,\phi}=\sum_{i=0}^{2J+1}c_{i}\left(\theta,\phi\right)\ket{J,J_{z}=J-i}
\end{align}
where $c_{i}\left(\theta,\phi\right)$ is correspond coefficients determined by rotating angle $\theta$ and $\phi$. In order to implement the spin coherent state on the cold atom system, one could map spin states into $2J+1$ focks states. For instance, one of maps is $\ket{J,J_{z}=J-i}\Longrightarrow\ket{i}$ so that the spin coherent state becomes
\begin{align}
    \ket{\theta,\phi}=\sum_{i=0}^{2J+1}c_{i}\left(\theta,\phi\right)\ket{i}
\end{align}
However, both spin $2$-cat code and $2$-cat code suffer from the boson loss too much. Thus, we just choose CLY code and binomial code for the further numerical simulation.

\end{document}